\documentstyle[11pt,aaspp4,flushrt]{article}

\newcommand{\apl}{\lesssim}
\newcommand{\cmjj}{\mbox{${\rm cm^{-2}}$}}
\newcommand{\etal}{et al.}

\newcommand{\ibid}{\underline{\makebox[0.5in]{}}.}

\newcommand{\kms}{\mbox{km\ s${^{-1}}$}}
\newcommand{\lya}{\mbox{${\rm Ly}\alpha$}}

\begin{document}

\lefthead{Chen et al.}

\righthead{Imaging of \lya-Absorbing Galaxies}

\title{THE GASEOUS EXTENT OF GALAXIES AND THE ORIGIN OF \lya\ ABSORPTION
SYSTEMS.  III.  HUBBLE SPACE TELESCOPE IMAGING OF \lya-ABSORBING GALAXIES AT $z
< 1$\altaffilmark{1}}
\altaffiltext{1}{Based on observations with the NASA/ESA Hubble Space
Telescope, obtained at the Space Telescope Science Institute, which is operated
by the Association of Universities for Research in Astronomy, Inc., under NASA
contract NAS5--26555.}

\author{HSIAO-WEN CHEN and KENNETH M. LANZETTA}
\affil{Department of Physics and Astronomy, State University of New York at
Stony Brook \\
Stony Brook, NY 11794--3800, U.S.A. \\
hchen,lanzetta@sbast3.ess.sunysb.edu}

\author{JOHN K. WEBB}
\affil{School of Physics, University of New South Wales \\
Sydney 2052, NSW, AUSTRALIA \\
jkw@edwin.phys.unsw.edu.au}

\and

\author{XAVIER BARCONS}
\affil{Instituto de F\'\i sica de Cantabria (Consejo Superior de
Investigaciones Cient\'\i ficas--- \\
Universidad de Cantabria), Facultad de Ciencias \\
39005 Santander, SPAIN \\
and \\
Institute of Astronomy, Madingley Road, Cambridge CB3 0HA, UK \\
barcons@ifca.unican.es}

\slugcomment{\it Astrophysical Journal Accepted}

\newpage

\begin{abstract}

  We present initial results of a program to obtain and analyze HST WFPC2
images of galaxies identified in an imaging and spectroscopic survey of faint
galaxies in fields of HST spectroscopic target QSOs.  We measure properties of
87 galaxies, of which 33 are associated with corresponding \lya\ absorption
systems and 24 do not produce corresponding \lya\ absorption lines to within
sensitive upper limits.  Considering only galaxy and absorber pairs that are
likely to be physically associated and excluding galaxy and absorber pairs
within 3000 \kms\ of the background QSOs leaves 26 galaxy and absorber pairs 
and seven galaxies that do not produce corresponding \lya\ absorption lines to
within sensitive upper limits. Redshifts of the galaxy and absorber pairs range
from 0.0750 to 0.8912 with a median of 0.3718, and impact parameter separations
of the galaxy and absorber pairs range from 12.4 to $157.4 \ h^{-1}$ kpc with a
median of $62.4 \ h^{-1}$ kpc.  The primary result of the analysis is that the
amount of gas encountered along the line of sight depends on the galaxy impact
parameter and $B$-band luminosity but does not depend strongly on the galaxy
average surface brightness, disk-to-bulge ratio, or redshift.  This result
confirms and improves upon the anti-correlation between \lya\ absorption
equivalent width and galaxy impact parameter found previously by Lanzetta et
al.\ (1995).  Spherical halos cannot be distinguished from flattened disks on
the basis of the current observations, and there is no evidence that galaxy
interactions play an important role in distributing tenuous gas around galaxies
in most cases. Galaxies might account for all \lya\ absorption systems with 
$W > 0.3$ \AA, but this depends on the unknown luminosity function and gaseous 
cross sections of low-luminosity galaxies as well as on the uncertainties of 
the observed number density of \lya\ absorption systems.

\end{abstract}


\newpage

\section{INTRODUCTION}

  Although it is generally believed that low-redshift \lya\ absorption systems
are  associated with galaxies and thus trace the large-scale structure of the
universe, the existence of a physical connection between individual galaxies 
and individual \lya\  absorption systems is still a matter of some debate 
(Morris et al.\ 1993; Lanzetta et al.\ 1995; Stocke et al.\ 1995; Bowen, 
Blades, \& Pettini 1996; Le Brun, Bergeron, \& Boiss\'e 1996; van Gorkom et 
al.\ 1996).   Over the past several years we have been conducting an imaging 
and spectroscopic survey of faint galaxies in fields of Hubble Space Telescope 
(HST) spectroscopic target QSOs (Lanzetta \etal\ 1995; Lanzetta, Webb, \& 
Barcons 1995, 1996, 1997a,b; Barcons, Lanzetta, \& Webb 1995).  The goal of the
survey is to establish the relationship between galaxies and \lya\ absorption 
systems by directly comparing galaxies and \lya\ absorption systems along 
common lines of sight.  One of the most striking results of the survey is that 
there exists a distinct anti-correlation between \lya\ absorption equivalent 
width and galaxy impact parameter.  In particular, galaxies at impact 
parameters less than $\approx 160 \ h^{-1}$ kpc are {\em almost always} 
associated with corresponding \lya\ absorption lines, whereas galaxies at 
impact parameters greater than $\approx 160 \ h^{-1}$ kpc are {\em almost 
never} associated with corresponding \lya\ absorption lines.  On the basis of 
this result, we conclude that most galaxies are surrounded by extended gaseous 
envelopes of $\approx 160 \ h^{-1}$ kpc radius and that many or most \lya\ 
absorption systems arise in extended gaseous envelopes of galaxies, where by 
``gaseous envelope'' we mean simply a gaseous structure of large covering 
factor but unspecified geometry or filling factor.

  But the scatter about the mean relationship between \lya\ absorption
equivalent width and galaxy impact parameter is substantial---specifically, the
dispersion spans roughly a factor of two.  Evidently the amount of gas
encountered along the line of sight depends on other factors besides galaxy
impact parameter, including perhaps galaxy luminosity, size, or morphological
type, geometry of the impact (e.g.\ if tenuous gas is distributed around
galaxies in flattened disks rather than in spherical halos), or disturbed
morphologies or the presence of close companions (e.g.\ if tenuous gas is
distributed around galaxies as a result of galaxy interactions).  To determine
these other factors, we have initiated a program to obtain and analyze HST Wide
Field Planetary Camera 2 (WFPC2) images of galaxies identified in the survey.

  Here we present initial results of this program, based on analysis of HST
images of six fields obtained in Cycle 5 and four fields accessed from the HST
archives.  From these images, we measure properties of 87 galaxies, of which 33
are associated with corresponding \lya\ absorption lines and 24 do not produce
corresponding \lya\ absorption lines to within sensitive upper limits. 
Considering only galaxy and absorber pairs that are likely to be physically
associated and excluding galaxy and absorber pairs within 3000 \kms\ of the
background QSOs leaves 26 galaxy and absorber pairs and seven galaxies that do
not produce corresponding \lya\ absorption lines to within sensitive upper
limits. Redshifts of the galaxy and absorber pairs range from 0.0750 to 0.8912
with a median of 0.3718, and impact parameter separations of the galaxy and
absorber pairs range from 12.4 to $157.4 \ h^{-1}$ kpc with a median of 
$62.4 \ h^{-1}$ kpc.

  The resulting measurements are used to address a variety of issues:  First,
the measurements are used to examine how the incidence and extent of tenuous 
gas around galaxies depends on galaxy luminosity, size, and morphological type.
Determining the fundamental scaling relationships between properties of the
galaxies and properties of the absorbers is the first step toward understanding
the nature and origin of the gas, analogous to establishing the Holmberg (1975)
and Bosma (1981) relationships for the outer, tenuous parts of galaxies. 
Second, the measurements are used to examine the rough geometry of tenuous gas 
around galaxies.  If galaxies are surrounded by spherical halos then they 
should exhibit absorption signatures that are independent of inclination and
orientation, whereas if galaxies are surrounded by flattened disks then they
should exhibit absorption preferentially at low inclination angles.  
Determining the rough distribution of tenuous gas around galaxies is crucial 
for distinguishing competing models of the origin of the gas.  And finally, the
measurements are used to examine the possibility that tenuous gas around
galaxies results from galaxy interactions, as evidenced by disturbed 
morphologies and the presence of close companions.

  The primary result of the analysis is that the amount of gas encountered
along the line of sight depends on the galaxy impact parameter and $B$-band
luminosity but does not depend strongly on the galaxy average surface
brightness, disk-to-bulge ratio, or redshift.  This result confirms and 
improves upon the anti-correlation between \lya\ absorption equivalent width 
and galaxy impact parameter found previously by Lanzetta et al.\ (1995) and 
argues that the gas is physically associated with the individual galaxies.  
Spherical halos cannot be distinguished from flattened disks on the basis of 
the current observations, and there is no evidence that galaxy interactions 
play an important role in distributing tenuous gas around galaxies in most 
cases.  A dimensionless Hubble constant $h = H_0/(100 \ {\rm km} \ {\rm s}^{-1}
\ {\rm Mpc}^{-1})$ and a deceleration parameter $q_0 = 0.5$ are adopted
throughout.

\section{OBSERVATIONS}

\subsection{WFPC2 Imaging Observations}

  Imaging observations of the fields surrounding 0349$-$1438, 0405$-$1219,
0850$+$4400, 1001$+$2910, 1354$+$1933, and 1704$+$6048 were obtained with HST
using WFPC2 with the F702W filter in Cycle 5.  The observations were obtained 
in a series of three or four exposures of 600 or 800 s each.  The journal of
observations is given in Table 1, which lists the field, 1950 coordinates
$\alpha$ and $\delta$ of the QSO, emission redshift $z_{\rm em}$ of the QSO,
filter, exposure time, and date of observation.

  Imaging observations of objects in the fields surrounding 0454$-$2203,
1545$+$2101, 1622$+$2352, and 2135$-$1446 were accessed from the HST archive.
The observations were obtained with the HST using WFPC2 with the F606W or F702W
filters.  The observations were obtained in a series of between two and 24
exposures of between 100 and 1000 s each.  The journal of archival observations
is given in Table 2, which lists the field, 1950 coordinates $\alpha$ and
$\delta$ of the QSO, emission redshift $z_{\rm em}$ of the QSO, filter, 
exposure time, and date of observation.

  The individual exposures were reduced using standard pipeline techniques and
were registered to a common origin, filtered for cosmic rays, and coadded using
our own reduction programs.  The final images are shown in Figure 1.  The
spatial resolution of the final images was measured to be ${\rm FWHM} \approx
0.1$ arcsec, and the $5 \sigma$ point source detection thresholds were measured
to span the range $m = 24.3$ through $m = 28.8$. 

\subsection{Other Observations}

  Detailed descriptions of the ground- and space-based imaging and
spectroscopic observations have been and will be presented elsewhere (e.g.\
Lanzetta et al.\ 1995), but in summary the observations consist of (1) optical
images and spectroscopy of objects in the fields of the QSOs, obtained with
various telescopes and from the literature, and (2) ultraviolet spectroscopy of
the QSOs, obtained with HST using the Faint Object Spectrograph (FOS) and
accessed through the HST archive.  These observations are summarized in Table
3, which for each field lists the number of galaxies (with spectroscopic
redshifts) included into the analysis, the reference to the galaxy observations
and analysis, the number of absorbers included into the analysis, and the
reference to the absorber observations and analysis.

\section{GALAXY IMAGE ANALYSIS}

  To determine structural parameters and angular inclinations and orientations
of the galaxies, we apply a surface brightness $\chi^2$ fitting program.  The
surface brightness $I(R)$ as a function of galactocentric radius $R$ is
represented as the sum of an azimuthally symmetric exponential disk component
and an azimuthally symmetric $R^{1/4}$ bulge component
\begin{equation}
I(R) = I_{0_D} \exp{\left\{ -1.68 \left[ \frac{R}{R_D} - 1 \right] \right\}} +
I_{0_B} \exp{\left\{ -7.67 \left[ \left( \frac{R}{R_B} \right)^{1/4} - 1
\right] \right\}},
\end{equation}
where $R_D$ and $R_B$ are the disk and bulge effective (or half-light) radii
and $I_{0_D}$ and $I_{0_B}$ are the disk and bulge effective surface
brightnesses (or surface brightnesses at the effective radii).  (The 
exponential scale length of the disk component is $R_D/1.68$.)  Galactocentric 
radius $R$ is related to projected radius $r$ and angle $\theta$ to the 
projected line segment joining the galaxy to the QSO for the disk component by
\begin{equation}
R = r \left[ 1 + \sin^2 (\theta + \alpha) \tan^2 i \right]^{1/2}
\end{equation}
and for the bulge component by
\begin{equation}
R = r \left[ 1 + \sin^2 (\theta + \alpha) \tan^2 (\arccos b/a) \right]^{1/2},
\end{equation}
where $\alpha$ is the orientation angle (measured north through east) between
the apparent major axis of the galaxy and the projected line segment joining 
the galaxy to the QSO and $i$ is the disk inclination angle and $b/a$ is the 
bulge axis ratio.  The theoretical image is determined by convolving the 
surface brightness distribution described by equations (1), (2), and (3) with 
a model of the HST point spread function obtained with the Tiny Tim program 
(Krist 1995). The estimates of $I_{0_D}$, $I_{0_B}$, $R_D$, $R_B$, $\alpha$,
$i$, and $b/a$ are determined by minimizing $\chi^2$ as calculated from the
theoretical image, the observed image, and the $1 \sigma$ error image, and the
uncertainties are calculated from the the Hessian matrix at the minimum.  The
disk-to-bulge ratio $D/B$ is the ratio of the integrated disk and bulge surface
brightness profiles.

  To determine the luminosities of the galaxies, we apply standard galaxy
photometry techniques.  The apparent magnitude $m$ (in AB magnitude units at 
the wavelength centroid of the filter response function) is determined by 
directly integrating the theoretical surface brightness distribution within the
Holmberg (1975) radius.  The $K$ correction is determined by integrating the 
product of the appropriate galaxy spectral energy distribution (selected by 
considering the galaxy spectral type) of Coleman, Wu, \& Weedman (1980) and the
appropriate filter response function and applying the appropriate bandpass 
correction.  The rest-frame $B$-band absolute magnitude $M_B - 5 \log h$ is 
determined from the apparent magnitude $m$ and the $K$ correction.  The 
rest-frame $B$-band luminosity of an $L_*$ galaxy (in $AB$ magnitude units at 
$B$) is taken to be $M_{B_*} = -19.5$ (Ellis et al.\ 1996). 

  To determine the rough morphological types of the galaxies, we consider the
visual appearances, surface brightness distributions, and disk-to-bulge ratios
$D/B$.  Generally, galaxies with $D/B < 3$ are classified as elliptical or S0
galaxies, galaxies with $3 < D/B < 14$ are classified as early-type spiral
galaxies, and galaxies with $D/B > 14$ are classified as late-type spiral
galaxies (Burstein 1979).

  The results are summarized in the ``Galaxies'' portion of Table 4, which for
each galaxy lists the field, Right Ascension and Declination offsets from the
QSO $\Delta \alpha$ and $\Delta \delta$, redshift $z_{\rm gal}$, impact
parameter $\rho$, disk and bulge effective radii $R_D$ and $R_B$, disk-to-bulge
ratio $D/B$, orientation angle $\alpha$, disk inclination angle $i$, bulge axis
ratio $b/a$, apparent magnitude $m$, average surface brightness $\langle \mu
\rangle$ (within the Holmberg radius), and absolute $B$-band magnitude 
$M_B - 5 \log h$.  Measurement uncertainties in $R_D$ and $R_B$ are typically 
2\%, measurement uncertainties in $D/B$ are typically 35\%, measurement 
uncertainties in $\alpha$ and $i$ are typically $2$ deg, and measurement 
uncertainties in $m$ and $M_B - 5 \log h$ are typically $0.2$.

\section{GALAXY AND ABSORBER PAIRS}

  Some galaxy and absorber pairs constitute {\em random} pairs, for which there
is no physical relationship between the galaxy and the absorber.  Random pairs
dominate at velocity and impact parameter separations large in comparison to 
the scale on which galaxies cluster.  Some galaxy and absorber pairs constitute
{\em correlated} pairs, for which the galaxy is correlated with the galaxy that
produces the absorber.  Correlated pairs dominate at velocity and impact
parameter separations comparable to the scale on which galaxies cluster but
large in comparison to the characteristic extent of tenuous gas  around
galaxies.  And some galaxy and absorber pairs constitute {\em physical} pairs,
for which the galaxy produces the absorber.  Physical pairs dominate at 
velocity and impact parameter separations comparable to the characteristic 
extent of tenuous gas around galaxies.  Because the goal of the analysis is to 
investigate tenuous gas around galaxies, it is necessary first to distinguish 
physical pairs from correlated and random pairs.

  Physical and correlated pairs are easily statistically distinguished from 
random pairs by means of the galaxy--absorber cross-correlation function 
$\xi_{\rm ga}(v,\rho)$.  Specifically, a galaxy and absorber pair of velocity 
separation $v$ and impact parameter separation $\rho$ is likely to be a 
correlated or physical pair if the cross-correlation amplitude satisfies 
$\xi_{\rm ga}(v,\rho) > 1$ and is likely to be a random pair if the 
cross-correlation amplitude satisfies $\xi_{\rm ga}(v,\rho) < 1$.  Physical 
pairs are less easily distinguished from correlated pairs, because both can 
occur at relatively small velocity and impact parameter separations.  To 
identify physical galaxy and absorber pairs we adopt the cross-correlation 
function $\xi_{\rm ga}(v,\rho)$ measured by Lanzetta, Webb, \& Barcons 
(1997a,b) on the basis of 352 galaxies and 230 absorbers in 24 fields, from 
which 3126 galaxy and absorber pairs are formed.  First, following Lanzetta et 
al.\ (1995) we accept \lya\ absorption lines according to a $3 \sigma$ 
detection threshold criterion, which is appropriate because the measurements 
are performed at a small number of known galaxy redshifts.  Next, we form 
galaxy and absorber pairs by requiring (1) $\xi_{\rm ga} > 1$ (which excludes 
likely random pairs) and (2) $\rho < 200 \ h^{-1}$ kpc (which from results of 
Lanzetta, Webb, \& Barcons 1997a,b excludes likely correlated pairs).  Next, 
we exclude galaxy and absorber pairs within 3000 \kms\ of the background QSOs 
(which are likely to be associated with the QSOs), and in cases where more 
than one galaxy is paired with one absorber we choose the galaxy at the 
smallest impact parameter.  Finally, we measure $3 \sigma$ upper limits to 
\lya\ absorption equivalent widths of galaxies that are not paired with 
corresponding absorbers, retaining only those measurements with $3 \sigma$ 
upper limits satisfying  $W < 0.35$ \AA.  This procedure identifies 26 galaxy 
and absorber pairs and seven galaxies that do not produce corresponding \lya\ 
absorption lines to within sensitive upper limits.  Redshifts of the galaxy 
and absorber pairs range from 0.0750 to 0.8912 with a median of 0.3718, and 
impact parameter separations of the galaxy and absorber pairs range from 12.4 
to $157.4 \ h^{-1}$ kpc with a median of $62.4 \ h^{-1}$ kpc.

  The results are summarized in the ``Absorbers'' portion of Table 4, which
for each galaxy lists the absorber redshift $z_{\rm abs}$ and \lya\ absorption
equivalent width $W$.  Measurement uncertainties in $W$ are typically 0.1 \AA.
In Table 4, galaxy entries without corresponding absorber entries represent
cases for which the absorption measurement cannot be made, either because the
galaxy occurs behind the QSO, the appropriate QSO spectrum is not available or
lacks sensitivity, the spectral region containing the predicted \lya\ line is
blended with other absorption lines, or a corresponding \lya\ absorption line
was paired with a galaxy at a smaller impact parameter.

  In adopting a quantitative criterion based on the galaxy--absorber
cross-correlation function to distinguish physical pairs from correlated and
random pairs, our analysis differs from the analyses of Le Brun, Bergeron, \&
Boiss\'e  (1996) and Bowen, Blades, \& Pettini (1996), which failed to detect
the anti-correlation between \lya\ absorption equivalent width and galaxy 
impact parameter.  These analyses included galaxy and absorber pairs at impact
parameter separations of up to $3800 \ h^{-1}$ kpc, of which approximately 65\%
are at separations of more than $200 \ h^{-1}$ kpc.  At these large impact
parameter separations, our evaluation using the galaxy--absorber
cross-correlation function indicates vanishingly small probabilities of finding
bona fide physical pairs.  We believe that this difference in the analyses
explains the differences in the results.

\section{DESCRIPTIONS OF INDIVIDUAL FIELDS}

  Here we present brief descriptions of the individual fields, noting galaxies
by their coordinate offsets in Right Ascension and Declination, respectively,
from the QSO line of sight in units of 0.1 arcsec.

\subsection{The Field toward 0349$-$1438}

  Galaxy $-00094$$+00108$ at $z = 0.3567$ and $\rho = 43.4 \ h^{-1}$ kpc is a 
late-type spiral galaxy of luminosity $L_B = 0.62 L_{B_*}$. This galaxy is
associated with a corresponding \lya\ absorption line at $z = 0.3566$ with
$W = 0.94$ \AA.

  Galaxies $+00117$$-00241$ at $z = 0.3236$ and $\rho = 77.0 \ h^{-1}$ kpc,
$-00292$$+00183$ at $z = 0.3244$ and $\rho = 99.1 \ h^{-1}$ kpc, and
$+00126$$-00459$ at $z = 0.2617$ and $\rho = 120.3 \ h^{-1}$ kpc are elliptical
or S0 galaxies of luminosities between $L_B = 0.30 L_{B_*}$ and $L_B = 0.97
L_{B_*}$.  Galaxy $-00571$$+00198$ at $z = 0.3273$ and $\rho = 174.7 \ h^{-1}$
kpc is a late-type spiral galaxy of luminosity $L_B = 0.73 L_{B_*}$. These
galaxies do not have sensitive \lya\ absorption measurements available.

\subsection{The Field toward 0405$-$1219}

  Galaxies $+00106$$-00067$ at $z = 0.5714$ and $\rho = 46.9 \ h^{-1}$ kpc,
$+00026$$-00150$ at $z = 0.5657$ and $\rho = 56.8 \ h^{-1}$ kpc,
$+00104$$-00269$ at $z = 0.5779$ and $\rho = 108.3 \ h^{-1}$ kpc,
$+00140$$-00354$ at $z = 0.5777$ and $\rho = 143.0 \ h^{-1}$ kpc, and
$-00369$$+00354$ at $z = 0.5696$ and $\rho = 191.1 \ h^{-1}$ kpc are elliptical
or S0 galaxies of luminosities between $L_B = 0.88 L_{B_*}$ and $L_B = 1.98 
L_{B_*}$. These galaxies (which occur in the immediate vicinity of the QSO) do 
not produce corresponding \lya\ absorption to within a sensitive upper limit.

  Galaxy $-00033$$-00127$ at $z = 0.5696$ and $\rho = 49.0 \ h^{-1}$ kpc shows 
a disturbed morphology, to which normal disk and bulge profiles cannot be 
applied. This galaxy (which occurs in the immediate vicinity of the QSO) does 
not produce corresponding \lya\ absorption to within a sensitive upper limit.

  Galaxy $+00010$$-00350$ at $z = 0.3520$ and $\rho = 105.4 \ h^{-1}$ kpc is a
late-type spiral galaxy of luminosity $L_B = 0.28 L_{B_*}$.  This galaxy is
associated with a corresponding \lya\ absorption line at $z = 0.3514$ with $W =
0.70$ \AA.

  Galaxy $-00662$$-00309$ at $z = 0.1525$ and $\rho = 126.3 \ h^{-1}$ kpc is a
late-type spiral galaxy of luminosity $L_B = 0.89 L_{B_*}$.  This galaxy is
associated with a corresponding \lya\ absorption line at $z = 0.1532$ with 
$W = 0.20$ \AA.

  Galaxy $-00372$$+00284$ at $z = 0.3617$ and $\rho = 142.8 \ h^{-1}$ kpc is an
elliptical or S0 galaxy of luminosity $L_B = 3.21 L_{B_*}$.  This galaxy is
associated with a corresponding \lya\ absorption line at $z = 0.3610$ with $W =
0.77$ \AA.

  Galaxy $+00161$$-00451$ at $z = 0.6167$ and $\rho = 183.9 \ h^{-1}$ kpc is a
late-type spiral galaxy of luminosity $L_B = 1.00 L_{B_*}$.  This galaxy occurs
behind the QSO and so is excluded from all analysis.

  Galaxy $+00319$$-00479$ at $z = 0.2973$ and $\rho = 157.4 \ h^{-1}$ kpc is an
elliptical or S0 galaxy of luminosity $L_B = 0.80 L_{B_*}$.  This galaxy is
associated with a corresponding \lya\ absorption line at $z = 0.2980$ with $W =
0.30$ \AA.

  Galaxy $-00656$$+00567$ at $z = 0.2800$ and $\rho = 228.7 \ h^{-1}$ kpc shows
signs of interaction and a disturbed morphology, to which normal disk and bulge
profiles cannot be applied.  This galaxy does not produce corresponding \lya\
absorption to within a sensitive upper limit.

  Galaxy $+00192$$-00700$ at $z = 0.5170$ and $\rho = 261.6 \ h^{-1}$ kpc is an
late-type spiral galaxy of luminosity $L_B = 0.70 L_{B_*}$.  This galaxy does
not produce corresponding \lya\ absorption to within a sensitive upper limit.

\subsection{The Field toward 0454$-$2203}

  Galaxy $+00106$$+00057$ at $z = 0.2784$ and $\rho = 31.6 \ h^{-1}$ kpc is a
late-type spiral galaxy of luminosity $L_B = 0.22 L_{B_*}$.  This galaxy is 
associated with a corresponding \lya\ absorption line at $z = 0.2777$ with 
$W = 0.81$ \AA.

  Galaxy $-00011$$-00180$ at $z = 0.4847$ and $\rho = 63.3 \ h^{-1}$ kpc is a
late-type spiral galaxy of luminosity $L_B = 2.12 L_{B_*}$, while galaxies
$-00182$$+00429$ at $z = 0.4836$ and $\rho = 163.4 \ h^{-1}$ kpc, and
$+00378$$-00365$ at $z = 0.4837$ and $\rho = 184.3 \ h^{-1}$ kpc are
spiral galaxies of luminosity $L_B \approx 1.9 L_{B_*}$.  These galaxies are 
associated with a corresponding \lya\ absorption line at $z = 0.4825$ with 
$W = 1.54$ \AA.  The small impact parameter of galaxy $-00011$$-00180$ 
suggests that it contributes most to the absorption line, although the modest 
impact parameters of galaxies $-00182$$+00429$ and $+00378$$-00365$ suggest 
that they also might contribute to the absorption line.

  Galaxy $+00042$$+00184$ at $z = 0.5325$ and $\rho = 68.8 \ h^{-1}$ kpc is an
elliptical or S0 galaxy of luminosity $L_B = 0.47 L_{B_*}$, and galaxy
$+00356$$-00410$ at $z = 0.5336$ and $\rho = 198.1 \ h^{-1}$ kpc is a late-type
spiral galaxy of luminosity $L_B = 1.05 L_{B_*}$.  These galaxies (which occur 
in the immediate vicinity of the QSO) do not produce corresponding \lya\ 
absorption to within sensitive upper limits.

  Galaxy $+00003$$-00197$ at $z = 0.3818$ and $\rho = 61.9 \ h^{-1}$ kpc is a
late-type spiral galaxy of luminosity $L_B = 0.64 L_{B_*}$.  This galaxy is
associated with a  corresponding \lya\ absorption line at $z = 0.3815$ with 
$W = 0.46$ \AA.  (A spiral arm of galaxy $-00011$$-00180$ points directly at 
galaxy $+00003$$-00197$, giving the impression that the two form an interacting
pair. But in fact the two occur at cosmologically distinct redshifts.)

  Galaxy $+00145$$+00995$ at $z = 0.3382$ and $\rho = 296.1 \ h^{-1}$ kpc is an
early-type spiral galaxy of luminosity $L_B = 0.76 L_{B_*}$.  This galaxy does 
not produce corresponding \lya\ absorption to within a sensitive upper limit.

\subsection{The Field toward 0850$+$4400}

  Galaxy $-00089$$+00020$ at $z = 0.1635$ and $\rho = 16.6 \ h^{-1}$ kpc is an
elliptical or S0 galaxy of luminosity $L_B = 0.40 L_{B_*}$.  This galaxy is
associated with a corresponding damped \lya\ absorption line at $z = 0.1630$
with $W = 5.9$ \AA.  This absorption system is the target of a recent
investigation by Lanzetta et al.\ (1997c), which on the basis of a
high-resolution spectrum obtained with HST using the Goddard High Resolution
Spectrograph establishes a neutral hydrogen column density of $\log N = 19.81
\pm 0.04$ \cmjj.

  Galaxy $+00100$$-00024$ at $z = 0.4402$ and $\rho = 34.6 \ h^{-1}$ kpc is an
S0 galaxy of luminosity $L_B = 0.80 L_{B_*}$ with visible faint disk.  This 
galaxy is associated with a corresponding \lya\ absorption line at $z = 0.4435$
with $W = 0.51$ \AA. 

  Galaxy $+00226$$+00124$ at $z = 0.5007$ and $\rho = 91.7 \ h^{-1}$ kpc is a
late-type spiral galaxy of luminosity $L_B = 1.05 L_{B_*}$.  Galaxy 
$+00455$$+00022$ at $z = 0.5196$ and $\rho = 164.5 \ h^{-1}$ kpc is an
elliptical or S0 galaxy of luminosity $L_B = 2.22 L_{B_*}$.  These galaxies 
(which occur in the immediate vicinity of the QSO) do not produce corresponding
\lya\ absorption to within a sensitive upper limit.

  Galaxy $+00290$$+00305$ at $z = 0.2766$ and $\rho = 110.2 \ h^{-1}$ kpc is a
late-type spiral galaxy of luminosity $L_B = 0.28 L_{B_*}$.  Galaxy 
$-00005$$+00405$ at $z = 0.0915$ and $\rho = 46.2 \ h^{-1}$ kpc is a
late-type spiral galaxy of luminosity $L_B = 0.07 L_{B_*}$.  Galaxy
$+00034$$+00350$ at $z = 0.0872$ and $\rho = 38.5 \ h^{-1}$ kpc is an
elliptical or S0 galaxy of luminosity $L_B = 0.34 L_{B_*}$.  These galaxies do
not have sensitive \lya\ absorption measurements available.

\subsection{The Field toward 1001$+$2910}

  Galaxy $-00034$$-00231$ at $z = 0.1380$ and $\rho = 37.6 \ h^{-1}$ kpc is a 
late-type spiral galaxy of luminosity $L_B = 0.03 L_{B_*}$.  This galaxy is
associated with a corresponding \lya\ absorption line at $z = 0.1377$ with
$W = 0.67$ \AA.

  Galaxy $-00284$$+00066$ at $z = 0.3308$ and $\rho = 85.0 \ h^{-1}$ kpc is an
elliptical or S0 galaxy of luminosity $L_B = 0.39 L_{B_*}$.  This galaxy (which
occurs in the immediate vicinity of the QSO) is associated with a corresponding
\lya\ absorption line at $z = 0.3283$ with $W = 0.18$ \AA.

\subsection{The Field toward 1354$+$1933}

  Galaxy $+00012$$+00075$ at $z = 0.4592$ and $\rho = 26.1 \ h^{-1}$ kpc is a 
late-type spiral galaxy of luminosity $L_B = 0.82 L_{B_*}$.  This galaxy is
associated with a corresponding \lya\ absorption line at $z = 0.4570$ with
$W = 1.41$ \AA.

  Galaxy $-00216$$-00122$ at $z = 0.4406$ and $\rho = 83.5 \ h^{-1}$ kpc is
an elliptical or S0 galaxy of luminosity $L_B = 0.65 L_{B_*}$.  This galaxy 
does not produce corresponding \lya\ absorption to within a sensitive upper 
limit.

  Galaxy $-00160$$-00233$ at $z = 0.4295$ and $\rho = 94.1 \ h^{-1}$ kpc is a 
late-type spiral galaxy of luminosity $L_B = 0.38 L_{B_*}$.  This galaxy is
associated with a corresponding \lya\ absorption line at $z = 0.4306$ with
$W = 1.03$ \AA.

  Galaxy $-00134$$-00480$ at $z = 0.5293$ and $\rho = 181.3 \ h^{-1}$ kpc is
an elliptical or S0 galaxy of luminosity $L_B = 0.91 L_{B_*}$.  This galaxy 
does not produce corresponding \lya\ absorption to within a sensitive upper 
limit.

  Galaxy $+00658$$+00501$ at $z = 0.3509$ and $\rho = 248.5 \ h^{-1}$ kpc is an
elliptical or S0 galaxy of luminosity $L_B = 0.23 L_{B_*}$.  This galaxy does
not produce corresponding \lya\ absorption to within a sensitive upper limit.

\subsection{The Field toward 1545$+$2101}

  Galaxy $-00027$$+00011$ at $z = 0.2657$ and $\rho = 7.2 \ h^{-1}$ kpc is an
elliptical or S0 galaxy of luminosity $L_B = 0.67 L_{B_*}$, and galaxy
$+00166$$-00087$ at $z = 0.2639$ and $\rho = 47.6$ is an elliptical or S0
galaxy of luminosity $L_B = 0.92 L_{B_*}$.  These galaxies (which occur in the
immediate vicinity of the QSO) are associated with a corresponding \lya\
absorption line at $z = 0.2641$ with $W = 0.63$.  The small impact parameters
of galaxies $-00027$$-00011$ and $+00166$$-00087$ suggest that both might
contribute to the absorption line, although spectroscopic observations of four
other galaxies toward 1545$+$2101 presented by Lanzetta, Webb, \& Barcons
(1996) indicate that these galaxies belong to a group or cluster of galaxies,
several or all of which might contribute to the \lya\ absorption line.

  Galaxy $-00263$$+00231$ at $z = 0.1343$ and $\rho = 55.1 \ h^{-1}$ kpc is a
late-type spiral galaxy of luminosity $L_B = 0.22 L_{B_*}$.  This galaxy does 
not produce corresponding \lya\ absorption to within a sensitive upper limit.  
This is particularly striking because almost all other galaxies yet identified 
at $\rho < 100 \ h^{-1}$ kpc are associated with corresponding \lya\ absorption
lines, except for  galaxies in the immediate vicinities of QSOs.

  Galaxy $+00273$$-$00356 at $z = 0.0949$ and $\rho = 53.4 \ h^{-1}$ kpc is a
late-type spiral galaxy of luminosity $L_B = 0.03 L_{B_*}$.  This galaxy is
associated with a corresponding \lya\ absorption line at $z = 0.0961$ with $W =
0.18$ \AA.

\subsection{The Field toward 1622$+$2352}

  The relationship between galaxies and absorption systems in this field was
analyzed previously by Steidel et al.\ (1997), concentrating on  metal-selected
absorption systems.  Our analysis differs from the analysis of Steidel et al.\
(1997) in two ways:  (1) we applied the surface brightness analysis of \S\ 3 to
all galaxies with identified redshifts and (2) we accepted \lya\ absorption
lines according to a $3 \sigma$ detection threshold criterion and measured $3
\sigma$ upper limits to \lya\ absorption equivalent widths of galaxies that are
not paired with corresponding absorbers.

  Galaxies $-00015$$+00021$ at $z = 0.931$ and $\rho = 10.9 \ h^{-1}$ kpc, 
$+00073$$-00121$ at $z = 0.920$ and $\rho = 59.5 \ h^{-1}$ kpc, 
$+00088$$-00248$ at $z = 0.919$ and $\rho = 111.0 \ h^{-1}$ kpc, and 
$-00289$$+00057$ at $z = 0.923$ and $\rho = 124.1 \ h^{-1}$ kpc are late-type 
spiral galaxies of luminosities in the range $L_B = 0.31 L_{B_*}$  to 
$3.04 L_{B_*}$.  Galaxies $-00051$$+00111$ at $z = 0.920$ and 
$\rho = 51.4 \ h^{-1}$ kpc, $-00013$$+00113$ at $z = 0.921$ and $\rho = 
47.9 \ h^{-1}$ kpc, and $-00264$$-00027$ at $z = 0.924$ and 
$\rho = 111.8 \ h^{-1}$ kpc  are elliptical or S0 galaxies of luminosities 
$L_B = 0.81 L_{B_*}$ and $3.38 L_{B_*}$, respectively.  These galaxies (which 
occur in the immediate  vicinity of the QSO) do not produce corresponding 
\lya\ absorption to within  a sensitive upper limit.

  Galaxy $+00030$$-00000$ at $z = 0.892$ and $\rho = 12.4 \ h^{-1}$ kpc is a
highly inclined late-type spiral galaxy of luminosity $L_B = 0.65 L_{B_*}$, 
and galaxy $-00076$$-00434$ at $z = 0.892$ and $\rho = 184.6 \ h^{-1}$ kpc 
is a late-type spiral galaxy of luminosity $L_B = 0.44 L_{B_*}$.  These two
galaxies are associated with a corresponding \lya\ absorption line at $z =
0.8909$ with $W = 2.71$ \AA. The small impact parameter of galaxy 
$+00030$$-00000$ suggests that it contributes most to the absorption line.

  Galaxy $-00042$$-00038$ at $z = 0.4720$ and $\rho = 19.5 \ h^{-1}$ kpc is a
late-type spiral galaxy of luminosity $L_B = 0.22 L_{B_*}$.  This galaxy is
associated with a corresponding \lya\  absorption line at $z  = 0.4716$ with 
$W = 0.91$ \AA.  This absorption system is noted as an upper-limit in the
sample of Steidel \etal\ (1996).

  Galaxy $-00055$$+00075$ at $z = 0.6352$ and $\rho = 36.3 \ h^{-1}$ kpc is a
late-type spiral galaxy of luminosity $L_B = 0.11 L_{B_*}$.  This galaxy is
associated with a corresponding \lya\  absorption line at $z  = 0.6359$ with $W
= 0.47$ \AA.

  Galaxy $-00089$$+00032$ at $z = 0.798$ and $\rho = 39.0 \ h^{-1}$ kpc is a
late-type spiral galaxy of luminosity $L_B = 1.05 L_{B_*}$.  This galaxy is 
associated with a corresponding \lya\ absorption line at $z = 0.7964$ with 
$W = 1.37$ \AA.

  Galaxy $-00033$$+00091$ at $z = 0.5650$ and $\rho = 36.0 \ h^{-1}$ kpc is an
elliptical or S0 galaxy of luminosity $L_B = 0.10 L_{B_*}$.  This galaxy does 
not produce corresponding \lya\ absorption to within  a sensitive upper limit.

  Galaxy $+00016$$+00142$ at $z = 0.656$ and $\rho = 55.9 \ h^{-1}$ kpc is a
late-type spiral galaxy of luminosity $L_B = 0.43 L_{B_*}$.  This galaxy is
associated with a corresponding \lya\ absorption line at $z = 0.6564$ with $W =
7.75$ \AA.  Although the \lya\ absorption equivalent width associated with this
galaxy appears to be particularly large given the relatively large impact
parameter of the galaxy, Steidel et al.\ (1996) were unable to identify another
galaxy at a smaller impact parameter despite an intensive search.

  Galaxy $-00123$$+00099$ at $z = 0.7016$ and $\rho = 73.4 \ h^{-1}$ kpc is a
late-type spiral galaxy of luminosity $L_B = 1.11 L_{B_*}$.  This galaxy is 
associated  with a corresponding \lya\ absorption line at $z = 0.7020$ with 
$W = 0.48$ \AA.

  Galaxy $+00079$$-00170$ at $z = 0.828$ and $\rho = 77.7 \ h^{-1}$ kpc is an
elliptical or S0 galaxy of luminosity $L_B = 0.35 L_{B_*}$.  This galaxy is 
associated  with a corresponding \lya\ absorption line at $z = 0.8273$ with 
$W = 1.06$ \AA.

  Galaxy $-00079$$+00307$ at $z = 0.7090$ and $\rho = 126.7 \ h^{-1}$ kpc is a
late-type spiral galaxy of luminosity $L_B = 0.16 L_{B_*}$.  This galaxy is 
associated with a corresponding \lya\ absorption line at $z = 0.7090$ with 
$W = 0.20$ \AA.  This absorption system is not contained in the sample of 
Steidel \etal\ (1996).

  Galaxies $-00111$$-00043$ at $z = 1.037$ and $\rho = 51.0 \ h^{-1}$ kpc,
$+00182$$+00002$ at $z = 1.010$ and $\rho = 77.7 \ h^{-1}$ kpc, 
$+00150$$+00196$ at $z = 1.018$ and $\rho = 105.2 \ h^{-1}$ kpc, and 
$-00343$$+00032$ at $z = 1.011$ and $\rho = 146.9 \ h^{-1}$ kpc show 
disturbed morphologies.  These galaxies occur behind the QSO and so are 
excluded from all analysis.

  Galaxies $-00067$$+00097$ at $z = 0.318$ and $\rho = 33.6 \ h^{-1}$ kpc, 
$-00213$$-00062$ at $z = 0.368$ and $\rho = 68.4 \ h^{-1}$ kpc, 
$-00241$$+00053$ at $z = 0.368$ and $\rho = 75.9 \ h^{-1}$ kpc, 
$+00255$$+00176$ at $z = 0.261$ and $\rho = 78.1 \ h^{-1}$ kpc, 
$+00127$$-00312$ at $z = 0.280$ and $\rho = 88.9 \ h^{-1}$ kpc, 
$-00454$$+00077$ at $z = 0.668$ and $\rho = 181.1 \ h^{-1}$ kpc,
and $-00468$$+00077$ at $z = 0.638$ and $\rho = 183.9 h^{-1}$ kpc  together 
show morphologies in the range elliptical or S0 galaxies to late-type spiral 
galaxies of luminosities in the range $L_B = 0.03 L_{B_*}$ to $1.35 L_{B_*}$.  
These galaxies do not have sensitive \lya\ absorption measurements available.

\subsection{The Field toward 1704$+$6048}

  Galaxy $+00081$$-00271$ at $z = 0.3615$ and $\rho = 86.4 \ h^{-1}$ kpc is a
late-type spiral galaxy of luminosity $L_B = 0.68 L_{B_*}$.  This galaxy (which
occurs at the immediate vicinity of the QSO) is associated with a corresponding
\lya\ absorption line at $z = 0.3621$ with $W = 0.67$ \AA.

  Galaxy $+00268$$-00352$ at $z = 0.3731$ and $\rho = 136.9 \ h^{-1}$ kpc is an
elliptical or S0 galaxy of luminosity $L_B = 1.32 L_{B_*}$.  This galaxy (which
occurs at the immediate vicinity of the QSO) is associated with a corresponding
\lya\ absorption line at $z = 0.3716$ with $W = 0.36$ \AA.

  Galaxy $-00510$$+00183$ at $z = 0.0921$ and $\rho = 62.2 \ h^{-1}$ kpc is an 
early-type spiral galaxy of luminosity $L_B = 0.19 L_{B_*}$.  This galaxy is
associated with a corresponding \lya\ absorption line at $z = 0.0920$ with
$W = 0.89$ \AA. This galaxy was studied previously by Barcons, Lanzetta, \&
Webb (1995).

  Galaxy $-00700$$+00133$ at $z = 0.2260$ and $\rho = 163.6 \ h^{-1}$ kpc is a
late-type spiral galaxy of luminosity $L_B = 0.14 L_{B_*}$.  This galaxy does
not produce corresponding \lya\ absorption to within a sensitive upper limit.

  Galaxy $+00729$$+00059$ at $z = 0.1877$ and $\rho = 147.4 \ h^{-1}$ kpc is an
elliptical or S0  galaxy of luminosity $L_B = 0.78 L_{B_*}$.  This galaxy is
associated with a corresponding \lya\ absorption line at $z = 0.1880$ with
$W = 0.43$ \AA.

  Galaxy $+00544$$+00102$ at $z = 0.4033$ and $\rho = 178.7 \ h^{-1}$ kpc is a
late-type spiral galaxy of luminosity $L_B = 0.35 L_{B_*}$.  This galaxy occurs
behind the QSO and so is excluded from all analysis.

  Galaxies $-00313$$-00094$ at $z = 0.3380$ and $\rho = 96.2 \ h^{-1}$ kpc and
$-00142$$-00304$ at $z = 0.0713$ and $\rho = 30.8 \ h^{-1}$ kpc are late-type
spiral galaxies of luminosities  $L_B = 0.25 L_{B_*}$ and $L_B = 0.007 L_{B_*}
$, respectively.  These galaxies do not have sensitive \lya\ absorption
measurements available.

\subsection{The Field toward 2135$-$1446}

  Galaxies $+00047$$-00026$ at $z = 0.1996$ and $\rho = 11.3 \ h^{-1}$ kpc,
$-00073$$+00157$ at $z = 0.2000$ and $\rho = 36.6 \ h^{-1}$ kpc, and
$+00186$$-00335$ at $z = 0.1991$ and $\rho = 80.8 \ h^{-1}$ kpc, are elliptical
or S0 galaxies of luminosities in the range $L_B = 0.39 L_{B_*}$ to $0.61
L_{B_*}$.  Galaxy $+00131$$-00217$ at $z = 0.1986$ and $\rho = 53.3 \ h^{-1}$ 
kpc and $+00284$$-00313$ at $z = 0.2011$ and $\rho = 89.6 \ h^{-1}$
kpc are  late-type spiral galaxies of luminosities $L_B = 0.25 L_{B_*}$ and
$L_B = 1.89 L_{B_*}$, respectively.  These galaxies (which occur in the 
immediate vicinity of the QSO) are associated with a corresponding \lya\ 
absorption line at $z = 0.2005$ with $W = 1.80$ \AA.  The small impact 
parameters of galaxies $+00047$$-00026$, $-00073$$+00157$, $+00131$$-00217$, 
$+00186$$-00335$, and $+00284$$-00313$ suggest that all might contribute to 
the absorption line.  Another galaxy just 1.9 arcsec from the QSO is visible 
in the image of $+00047$$-00026$.  If this galaxy is at the redshift
$z \approx 0.20$ of the other galaxies, it occurs at an impact parameter of 
only $\rho = 4.0 \ h^{-1}$ kpc.

  Galaxy $-00008$$+00493$ at $z = 0.0752$ and $\rho = 47.8 \ h^{-1}$ kpc is a
late-type spiral galaxy of luminosity $L_B = 0.27 L_{B_*}$.  This galaxy is 
associated with a corresponding \lya\ absorption line at $z = 0.0750$ with 
$W = 0.33$ \AA.  This galaxy was studied previously by Barcons, Lanzetta, \&
Webb (1995).

  Galaxy $+00508$$-00217$ at $z = 0.1857$ and $\rho = 111.0 \ h^{-1}$ kpc is a
late-type spiral galaxy of luminosity $L_B = 0.41 L_{B_*}$.  This galaxy is
associated with a corresponding \lya\ absorption line at $z = 0.1848$ with $W
= 1.29$ \AA.

\section{ANALYSIS}

\subsection{Method of Analysis}

  The goals of the analysis are (1) to demonstrate the existence of a fiducial
relationship between some measure of the strength of neutral hydrogen
absorption (e.g.\ \lya\ absorption equivalent width $W$ or neutral hydrogen
column density $N$) and galaxy impact parameter and (2) to assess whether
accounting for measurements of other galaxy properties (besides galaxy impact
parameter) can improve upon the fiducial relationship.  Conceptually, we divide
the the available measurements into a ``dependent measurement'' $y$ of the
strength of neutral hydrogen absorption and various ``independent 
measurements'' $x_1, x_2, ...$ of galaxy properties.  For the dependent 
measurement we consider variously $\log W$, $\log (W \cos i)$, $\log N$, and 
$\log (N \cos i)$.  (The $\cos i$ factor is included as a path-length 
correction in models of inclined galaxy disks.)   For the independent 
measurements we consider variously combinations of $\log \rho$, $\log L_B$, 
$\log r_e$, $\langle \mu \rangle$, $\log D/B$, and $\log (1+z)$.  The goals 
of the analysis are equivalent to searching for a ``fundamental surface'' in
a multi-dimensional space spanned by the dependent measurements and various 
combinations of the independent measurements.

  To determine the functional form of the relationship between the dependent 
and the independent measurements, we adopt a parameterized functional 
dependence
\begin{equation}
y = y(x_1, x_2, ...)
\end{equation}
and apply a maximum-likelihood analysis to estimate the parameters and their
uncertainties.   In practice, we consider the independent measurements one or
two at a time and adopt a linear or bi-linear relationship between the
dependent and the independent measurements
\begin{equation}
y = a_1 x_1 + {\rm constant}
\end{equation}
or
\begin{equation}
y = a_1 x_1 + a_2 x_2 + {\rm constant},
\end{equation}
where $x_1$ is always $\log \rho$ and $x_2$ is some other galaxy property.
(Except for the independent measurements $\langle \mu \rangle$, the
dependent and the independent measurements are always logarithms of actual
observed quantities, so in these cases equations 5 and 6 are equivalent to
power-law relationships between the actual observed quantities.)  Because the
observations include both measurements and upper limits, the likelihood 
function is taken to be
\begin{equation}
{\cal L} = \left( \prod_{i=1}^{n} \exp \left\{ -\frac{1}{2} \left[ \frac{y_i -
y(x_{1_i}, x_{2_i})}{\sigma_i} \right]^2 \right\} \right) \left(
\prod_{i=1}^m \int_{y_i}^{-\infty} dy' \exp \left\{ -\frac{1}{2} \left[ 
\frac{y' - y(x_{1_i}, x_{2_i})}{\sigma_i} \right]^2 \right\} \right),
\end{equation}
where $\sigma_i$ is the uncertainty of the dependent measurement and where the
first product extends over the $n$ measurements and the second product extends
over the $m$ upper limits.  (This definition of the likelihood function is
appropriate if the residuals about the mean relationship are normally
distributed.)  Because $\sigma_i$ may include significant ``cosmic'' scatter
(which presumably arises due to intrinsic variations between individual
galaxies) as well as measurement error, $\sigma_i$ is taken to be the
quadratic sum of the cosmic scatter $\sigma_c$ and the measurement error
$\sigma_{m_i}$
\begin{equation}
\sigma_i^2 = \sigma_c^2 + \sigma_{m_i}^2,
\end{equation}
where the cosmic scatter is defined by
\begin{equation}
\sigma_c^2 =  {\rm med} \left( \left\{ y_i - y(x_{1_i},x_{2_i}) -
\frac{1}{n} \sum_{j=1}^n \left[ y_j - y(x_{1_j},x_{2_j}) \right]
\right\}^2 - \sigma_{m_i}^2 \right).
\end{equation}
Because $\sigma_c$ depends on the maximum-likelihood solution
$y(x_{1_i},x_{2_i})$, the maximum-likelihood solution is obtained iteratively
with respect to equations (7) and (9).

  To assess the statistical significance of the relationship between the
dependent and the independent measurements, we apply two statistical tests of
the null hypothesis that the dependent and the independent measurements are
unrelated.  First, we apply a ``confidence interval test'' by using the
likelihood function to evaluate the $1 \sigma$ confidence interval on the
parameters.  Under the null hypothesis, $a_1 = a_2 = 0$ to within measurement
error, so a significant measurement of $a_1 \neq 0$ or $a_2 \neq 0$ invalidates
the null hypothesis.  Second, we apply an ``anti-correlation test'' by using 
the generalized (to treat upper limits) Kendall, Kendall, Spearman, and Pearson
correlation tests to evaluate the correlation between the dependent 
measurements and a ``reduced'' combination of the independent measurements 
$x_1 + (a_2/a_1) x_2$. (In applying the Kendall, Spearman, and Pearson 
correlation tests, which are not designed to treat upper limits, $3 \sigma$ 
upper limits are included as detections.)  Under the null hypothesis the 
dependent and independent measurements are uncorrelated, so a significant 
measurement of a correlation invalidates the null hypothesis.  Results of the 
statistical tests are presented in Table 5, which lists the measurements; the 
statistical significances $a/\delta a$ of the fitting coefficients; the 
correlation coefficients $r_{gk}$, $r_k$, $r_s$, and $r_p$ and statistical 
significances $r_{gk}/\sigma_{gk}$, $r_k/\sigma_k$, $r_s/\sigma_s$, and 
$r_p/\sigma_p$ of the generalized Kendall, Kendall, Spearman, and Pearson 
correlation tests, respectively; and the cosmic scatter $\sigma_c$.

\subsection{Galaxy Impact Parameter}

  In this section, we assess the possibility that the amount of gas intercepted
along the line of sight depends on galaxy impact parameter $\rho$ by adopting a
power-law relationship between $W$ and $\rho$
\begin{equation}
\log W = -\alpha \log \rho + {\rm constant}
\end{equation}
and applying the analysis of \S\ 6.1 to the measurements of \S\ 3.  The
maximum-likelihood analysis yields
\begin{equation}
\alpha = 0.93 \pm 0.13,
\end{equation}
and the results of the statistical tests are given in row 1 of Table 5.  The
relationship between $W$ and $\rho$ is shown in Figure 2.

  The results summarized in Table 5 indicate that $W$ is anti-correlated with
$\rho$.  Specifically, the confidence interval test indicates that $\alpha$
differs from zero at the $7.2 \sigma$ level of significance, and the
anti-correlation test indicates that $W$ is anti-correlated with $\rho$ at a
level of significance ranging from $3.36 \sigma$ through $4.45 \sigma$.  We
conclude that the amount of gas intercepted along the line of sight depends on
galaxy impact parameter, which argues that the gas is physically associated
with the individual galaxies.

\subsection{Galaxy $B$-Band Luminosity}

  In this section, we assess the possibility that the the amount of gas
intercepted along the line of sight depends on galaxy $B$-band luminosity
$L_B$ by adopting a power-law relationship between $W$ and $\rho$ and $L_B$
\begin{equation}
\log W = -\alpha \log \rho + \beta \log L_B + {\rm constant}
\end{equation}
and applying the analysis of \S\ 6.1 to the measurements of \S\ 3.  The
maximum-likelihood analysis yields
\begin{equation}
\alpha = 1.02 \pm 0.12
\end{equation}
and
\begin{equation}
\beta = 0.37 \pm 0.10,
\end{equation}
and the results of the statistical tests are given in row 2 of Table 5.  The
relationship between $W$ and $\rho$ and $L_B$ is shown in Figure 3.

  The results summarized in Table 5 indicate that the relationship between $W$
and $\rho$ accounting for $L_B$ is superior to the fiducial relationship 
between $W$ and $\rho$.  Specifically, the confidence interval test indicates 
that $\beta$ differs from zero at the $3.7 \sigma$ level of significance, and 
the anti-correlation test indicates that the anti-correlation between $W$ and 
$\rho$ accounting for $L_B$ is stronger and more significant than the 
anti-correlation between $W$ and $\rho$, at a level of significance ranging 
from $4.26 \sigma$ to $5.79 \sigma$.  We conclude that the amount of gas 
intercepted along the line of sight depends on galaxy $B$-band luminosity.  
This result applies over the $B$-band luminosity interval $0.007 L_{B_*} \apl 
L_B \apl 3.4 L_{B_*}$ spanned by the observations.

\subsection{Galaxy Effective Radius}

  In this section, we assess the possibility that the amount of gas intercepted
along the line of sight depends on galaxy effective radius $r_e$ by adopting a
power-law relationship between $W$ and $\rho$ and $r_e$
\begin{equation}
\log W = -\alpha \log \rho + \eta \log r_e + {\rm constant}
\end{equation}
and applying the analysis of \S\ 6.1 to the measurements of \S\ 3.  The results
of the statistical tests are given in row 3 of Table 5.

  The results summarized in Table 5 indicate that the relationship between $W$
and $\rho$ accounting for $r_e$ is superior to the fiducial relationship 
between $W$ and $\rho$ but marginally inferior to the relationship between $W$ 
and $\rho$ accounting for $L_B$.  Specifically, the confidence interval test
indicates that $\eta$ differs from zero at the $3.7 \sigma$ level of
significance, and the anti-correlation test indicates that the anti-correlation
between $W$ and $\rho$ accounting for $r_e$ is stronger and more
significant than the anti-correlation between $W$ and $\rho$ although neither
as strong nor as significant than the anti-correlation between $W$ and $\rho$
accounting for $L_B$.  The similarity between results accounting for $L_B$ and
results accounting for $r_e$ is plausible given the Holmberg (1975) correlation
between galaxy luminosity and size.  We conclude that the amount of gas
intercepted along the line of sight depends on galaxy effective radius, but
only indirectly through the correlation between galaxy effective radius and
galaxy $B$-band luminosity.  This result applies over the effective radius
interval $0.2 \apl r_e \apl 7.1 \ h^{-1}$ kpc spanned by the observations.

\subsection{Galaxy Average Surface Brightness}

  In this section, we assess the possibility that the amount of gas intercepted
along the line of sight depends on galaxy average surface brightness $\langle
\mu \rangle$ by adopting a power-law relationship between $W$ and $\rho$ and
$\langle \mu \rangle$
\begin{equation}
\log W = -\alpha \log \rho + \lambda \langle \mu \rangle + {\rm constant}.
\end{equation}
and again applying the analysis of \S\ 6.1 to the measurements of \S\ 3.  The
results of the statistical tests are given in row 4 of Table 5.

  The results summarized in Table 5 indicate that the relationship between $W$
and $\rho$ accounting for $\langle \mu \rangle$ is statistically identical to
the fiducial relationship between $W$ and $\rho$.  We conclude that the amount
of gas intercepted along the line of sight does not depend strongly on galaxy
average surface brightness.  This result applies over the average surface
brightness interval $21.3 \apl \langle \mu \rangle \apl 26.4$ mag arcsec$^{-2}$
spanned by the observations.

\subsection{Galaxy Disk-to-Bulge Ratio}

  In this section, we assess the possibility that the amount of gas intercepted
along the line of sight depends on galaxy disk-to-bulge ratio $D/B$ by adopting
a power-law relationship between $W$ and $\rho$ and $D/B$
\begin{equation}
\log W = -\alpha \log \rho + \delta \log D/B + {\rm constant}.
\end{equation}
and again applying the analysis of \S\ 6.2 to the measurements of \S\ 3.  The
results of the statistical tests are given in row 5 of Table 5.

  The results summarized in Table 5 indicate that the relationship between $W$
and $\rho$ accounting for $D/B$ is marginally superior to the fiducial
relationship between $W$ and $\rho$ but marginally inferior to the relationship
between $W$ and $\rho$ accounting for $L_B$.  We conclude that the amount of 
gas intercepted along the line of sight does not depend strongly on galaxy
disk-to-bulge ratio.  This result applies over the disk-to-bulge ratio interval
of bulge-dominated to disk-dominated galaxies spanned by the observations.

\subsection{Redshift}

  In this section, we assess the possibility that the amount of gas intercepted
along the line of sight depends on redshift $z$ by adopting a power-law
relationship between $W$ and $\rho$ and $1 + z$
\begin{equation}
\log W = -\alpha \log \rho + \gamma \log (1 + z) + {\rm constant}.
\end{equation}
and again applying the analysis of \S\ 6.2 to the measurements of \S\ 3.  The
results of the statistical tests are given in row 6 of Table 5.

  The results summarized in Table 5 indicate that the relationship between $W$
and $\rho$ accounting for $z$ is statistically identical to the fiducial
relationship between $W$ and $\rho$.  We conclude that the amount of gas
intercepted along the line of sight does not depend strongly on redshift.  This
result applies over the redshift interval $0.1 \apl z \apl 0.8$ spanned by the
observations.

\subsection{Shape of Tenuous Gas around Galaxies}

  In this section, we assess the possibility that tenuous gas is distributed
around galaxies in flattened disks rather than in spherical halos by repeating
the analyses of \S\S\ 6.2 through 6.7, substituting $W$ or $W \cos i$ for $W$
and galaxy galactocentric distance $R$ for galaxy impact parameter $\rho$.  If
tenuous gas is distributed around galaxies in flattened disks, then the 
strength of neutral hydrogen absorption should vary as a function of $R$ rather
than $\rho$.  (Galactocentric radius $R$ is related to impact parameter $\rho$
through the inclination and orientation angles $i$ and $\alpha$.)  In the
optically thick limit, the \lya\ absorption equivalent width should be
independent of path length through the disk, hence the \lya\ absorption
equivalent width corrected to face-on inclination is $W$.  In the optically 
thin limit, the \lya\ absorption equivalent width should vary as the path 
length through the disk, or in other words as $1/\cos i$, hence the \lya\ 
absorption equivalent width corrected to face-on inclination is $W \cos i$.  
The results of the statistical tests are given in rows 7 through 13 of Table 5,
and the relationship between $W \cos i$ and $R$ is shown in Figure 4.

  The results summarized in Table 5 indicate that the relationships between $W$
and $R$ or $W \cos i$ and $R$ are statistically identical to the fiducial
relationship between $W$ and $\rho$.  We conclude that spherical halos cannot 
be distinguished from flattened disks on the basis of the current observations.

\subsection{Column Density Distribution of Tenuous Gas around Galaxies}

  In this section, we assess the possibility that tenuous gas around galaxies
is better described by a relationship in neutral hydrogen column density $N$
than by a relationship in \lya\ absorption equivalent width by repeating the
analyses of \S\S\ 6.2 through 6.8 substituting $N$ or $N \cos i$ for $W$ or $W
\cos i$.  Following Lanzetta et al.\ (1995), we estimate neutral hydrogen 
column densities from \lya\ absorption equivalent widths by applying a 
curve-of-growth analysis, assuming that the Doppler parameters are contained 
in the range $20 < b < 40$ \kms.  The results of the statistical tests are 
given in rows 14 through 26 of Table 5.

  The results summarized in Table 5 indicate that the relationship between $N$
and $\rho$ or $N$ and $R$ are marginally superior to the relationships between
$W$ and $\rho$ or $W$ and $R$.  We conclude that tenuous gas around galaxies is
marginally better described by a relationship in neutral hydrogen column 
density than by a relationship in \lya\ absorption equivalent width.

  As a summary of the results of the preceding sections, we derive the column
density distribution of tenuous gas around galaxies, incorporating significant
scaling relationships and rejecting insignificant scaling relationships.  The
results of \S\S\ 6.2 through 6.8 indicate that neutral hydrogen column density
depends on $\rho$ and $L_B$ but does not depend strongly on $\langle \mu
\rangle$, $D/B$, or $z$ and that spherical halos cannot be distinguished from
flattened disks on the basis of the current observations.  Accordingly we adopt
a power-law relationship between $N$ and $\rho$ and $L_B$
\begin{equation}
\log \left( \frac{N}{10^{20} \ \cmjj} \right)  = -\alpha \log \left(
\frac{\rho}{10 \ {\rm kpc}} \right) + \beta \log \left( \frac{L_B}{L_{B_*}}
\right) + {\rm constant}
\end{equation}
and apply the analysis of \S\ 6.1 to the measurements of \S\ 3.  The
maximum-likelihood analysis yields
\begin{equation}
\alpha = 5.33 \pm 0.50,
\end{equation}
\begin{equation}
\beta = 2.19 \pm 0.55,
\end{equation}
and
\begin{equation}
{\rm constant} = 1.09 \pm 0.90.
\end{equation}
The cosmic scatter about this relationship is
\begin{equation}
\sigma_c = 0.815,
\end{equation}
which corresponds to a factor of $\approx 6.5$ in neutral hydrogen column
density.  Substituting values from equations (20), (21), and (22) into
equation (19), our current best estimate of the column density distribution of
tenuous gas around galaxies is
\begin{equation}
\log \left( \frac{N}{10^{20} \ \cmjj} \right)  = -5.33 \; \log \left(
\frac{\rho}{10 \ {\rm kpc}} \right) - 2.19 \; \log \left( \frac{L_B}{L_{B_*}}
\right) + 1.09.
\end{equation}
The relationship between $N$ and $\rho$ and $L_B$ is shown in Figure 5.

\subsection{Galaxy Morphology}

  The analyses of \S\S\ 6.5 and 6.6 indicate that the amount of gas intercepted
along the line of sight does not depend strongly on galaxy surface brightness 
or galaxy disk-to-bulge ratio.  Because both galaxy surface brightness and 
galaxy disk-to-bulge ratio are correlated with morphological type, this implies
that the amount of gas intercepted along the line of sight does not depend 
strongly on morphological type.  Furthermore, the morphological types of 
galaxies that produce corresponding \lya\ absorption lines are directly 
observed to range from elliptical or S0 galaxies (e.g.\ galaxy $-00284$$+00066$
toward 1001$+$2910) through early-type  spiral galaxies (e.g.\ galaxy $+00508$$
-00217$ toward 2135$-$1446) through late-type spiral galaxies (e.g.\ galaxy 
$+00273$$-00356$ toward 1545$+$2101).  We conclude that galaxies that produce 
\lya\ absorption systems span a wide range of morphological types.  

\subsection{Galaxy Interactions}

  Of the galaxies shown in Figure 1, only galaxies $-00033$$-00127$ and
$-00656$$+00567$ toward 0405$-$1219 appear to exhibit obvious signs of 
disturbed morphologies.  Galaxy $-00033$$-00127$ occurs in the immediate
vicinity of the QSO and does not produce a corresponding \lya\ absorption line
to within a sensitive upper limit.  Galaxy $-00656$$+00567$ appears to be in a
``post-merger'' stage with a visible asymmetric ring surrounding the nucleus. 
This galaxy does not produce a corresponding \lya\ absorption lines to within a
sensitive upper limit, although it occurs at a relatively large impact 
parameter ($\rho = 228.7 \ h^{-1}$ kpc).  We conclude that there is no evidence
that tenuous gas is distributed around galaxies as a result of galaxy 
interactions in most cases, although it is of course possible that tenuous gas 
is distributed around galaxies as a result of galaxy interactions in some 
cases.

\section{DISCUSSION}

  The primary result of the analysis is that the amount of gas encountered 
along the line of sight depends on the galaxy impact parameter $\rho$ and 
$B$-band luminosity $L_B$ but does not depend strongly on the galaxy average 
surface brightness $\langle \mu \rangle$, disk-to-bulge ratio $D/B$, or 
redshift $z$.  Apparently extended gaseous envelopes are a common and generic 
feature of galaxies of a wide range of luminosity and morphological type, and 
\lya\ absorption systems trace a significant and representative portion of the 
galaxy population.  The most important implication of this result is that it 
provides for the first time a means of quantitatively relating statistical 
properties of \lya\ absorption systems to statistical properties of faint 
galaxies.

  First, we determine the relationship between galaxy gaseous radius $r$ and
galaxy $B$-band luminosity $L_B$, which is analogous to the Holmberg (1975)
relationship between galaxy stellar radius and galaxy $B$-band luminosity. 
Adopting the power-law relationship between $W$ and $\rho$ and $L_B$ of 
equation (12), the relationship between $r$ and $L_B$ at a fixed \lya\ 
absorption equivalent width threshold is
\begin{equation}
\frac{r}{r_*} = \left( \frac{L_B}{L_{B_*}} \right)^t,
\end{equation}
where $r_*$ is the gaseous radius of an $L_*$ galaxy and where $t = \beta /
\alpha$.  Adopting $\alpha$ and $\beta$ from equations (13) and (14) yields
\begin{equation}
t = 0.37 \pm 0.11,
\end{equation}
and adopting the maximum-likelihood analysis value of the constant of equation
(12) yields
\begin{equation}
r_* = 174_{-25}^{+38} \ h^{-1} \ {\rm kpc}
\end{equation}
for a \lya\ absorption equivalent width threshold $W = 0.3$ \AA.  This result
applies over the $B$-band luminosity interval $0.007 \apl L_B \apl 3.4 L_{B_*}$
spanned by the observations.  The value of $r_*$ that we derive here agrees
completely with previous results of Lanzetta \etal\ (1995).

  Next, we apply the relationship between galaxy gaseous radius and galaxy
$B$-band luminosity to predict the incidence of \lya\ absorption systems
produced by galaxies drawn from a known galaxy luminosity function. Integrating
over galaxies of $B$-band luminosity $L_B > L_{B_{\rm min}}$, the predicted
number density per line of sight $n(z)$ of \lya\ absorption systems is
\begin{equation}
n(z) = \frac{c}{H_0} (1 + z) (1 + 2 q_0 z)^{-1/2} \int^\infty_{L_{B_{\rm min}}}
dL_B \ \Phi(L_B,z) \pi r^2(L_B) \kappa \epsilon,
\end{equation}
where $c$ is the speed of light, $\Phi(L_B,z)$ is the galaxy luminosity
function, $\kappa$ is the halo covering factor (averaged over all inclination
angles), and $\epsilon$ is the fraction of galaxies that produce corresponding
\lya\ absorption systems.  Applying equation (25) and adopting a Schechter
(1976) galaxy luminosity function $\Phi(L_B) = \Phi_* (L_B/L_{B_*})^{-s}
\exp(-L_B/L_{B_*})$, the predicted number density per line of sight of \lya\
absorption absorption systems is
\begin{equation}
n(z) = \frac{c}{H_0} (1 + z) (1 + 2 q_0 z)^{-1/2} \pi r_*^2 \kappa \epsilon
\Phi_* \Gamma ( 1 + 2 t - s, L_{B_{\rm min}}/L_{B_*}),
\end{equation}
where $\Gamma$ is the incomplete gamma function.

  To evaluate equation (29), we take $\kappa = \epsilon = 1$ (which is
consistent with the results of \S\ 6), adopt values of $t$ and $r_*$ from
equations (26) and (27) (which is appropriate for absorption systems with \lya\
equivalent width satisfying $W > 0.3$ \AA), adopt $L_{B_{\rm min}} = 0.005
L_{B_*}$ (which is the minimum galaxy $B$-band luminosity at which equation 25
has been established), and consider several recent determinations of the galaxy
luminosity function at moderate redshifts.  Adopting the galaxy luminosity
function of Ellis et al.\ (1996) (which is characterized by $\Phi_* =
0.0148_{-0.0019}^{+0.0030} \ h^3$ Mpc$^{-3}$ and $s = 1.41_{-0.07}^{+0.12}$ 
over the redshift interval $0.15 < z < 0.35$ and includes low surface
brightness galaxies), the predicted number density of \lya\ absorption systems
with $W > 0.3$ \AA\ is $n = 10.4_{-5.3}^{+8.1}$.  Adopting the galaxy 
luminosity function of Zucca et al.\ (1997) (which is characterized by $\Phi_* 
= 0.020 \pm 0.004 \ h^3$ Mpc$^{-3}$ and $s = 1.22_{-0.07}^{+0.06}$ over the 
redshift interval $z < 0.3$), the predicted number density of \lya\ absorption 
systems with $W > 0.3$ \AA\ is $n = 10.3_{-5.0}^{+6.4}$.  And adopting the 
galaxy luminosity function of Lilly et al.\ (1995) (which is characterized by 
$\Phi_* = 0.0272 \ h^3$ Mpc$^{-3}$ and $s = 1.03$ over the redshift interval 
$0.2 < z < 0.5$), the predicted number density of \lya\ absorption absorption 
systems with $W > 0.3$ \AA\ is $n = 11.2$.  In comparison, the observed number 
density of \lya\ absorption systems with $W > 0.3$ \AA\ is $n = 21.7\pm8.5$ at
$z = 0.35$ (Bahcall et al.\ 1996).  Apparently, luminous galaxies can explain
about 50\% of \lya\ absorption systems with $W > 0.3$ \AA.

  But the predicted number density of \lya\ absorption systems depends strongly
on the chosen galaxy luminosity function, which inherits two well-known
uncertainties:  the excess of faint galaxies and the normalization.  In
addition, the analysis has so far neglected galaxies of luminosity $L_B < 0.007
L_{B_*}$.  If we adopt the galaxy luminosity function of Ellis \etal\ (1996) 
over the redshift interval $0.35 < z < 0.75$ (which is characterized by $\Phi_* 
= 0.0355_{-0.0200}^{+0.0291} \ h^3$ Mpc$^{-3}$ and $s = 1.45_{-0.18}^{+0.16}$), 
the predicted number density of \lya\ absorption systems with $W > 0.3$ \AA\ is 
$n = 30.6_{-27.4}^{+45.2}$ for $L_{B_{\rm min}} = 0.001 L_{B_*}$. If we adopt 
the modified luminosity function of Zucca \etal\ (1996) over the redshift 
interval $z < 0.3$ (which is characterized by a Schechter form with $\Phi_* = 
0.021 \ h^3$ Mpc$^{-3}$ and $s = 1.16$ for $M_B<-16.99$ and by a power-law form 
$(L/L_*)^{-u}$ with $u = 1.57$ for $M_B>-16.99$), the predicted number density 
of \lya\ absorption systems with $W > 0.3$ \AA\ is $n = 12.9$ for 
$L_{B_{\rm min}} = 0.001 L_{B_*}$.  This number is still insufficient to account
for all of the observed \lya\ absorption systems, but it shows that the faint-end
slope of the galaxy luminosity function has a significant influence in the
determination of the predicted number density of \lya\ absorption systems.  
We conclude that galaxies might account for all \lya\ absorption systems with 
$W > 0.3$ \AA, but this depends on the unknown luminosity function and gaseous 
cross sections of low-luminosity galaxies as well as on the uncertainties of the 
observed number density of \lya\ absorption systems.

\section{SUMMARY AND CONCLUSIONS}

  We present initial results of a program to obtain and analyze HST WFPC2
images of galaxies identified in an imaging and spectroscopic survey of faint
galaxies in fields of HST spectroscopic target QSOs.  We measure properties of
87 galaxies, of which 33 are associated with corresponding \lya\ absorption
systems and 24 do not produce corresponding \lya\ absorption lines to within
sensitive upper limits.  Considering only galaxy and absorber pairs that are
likely to be physically associated and excluding galaxy and absorber pairs
within 3000 \kms\ of the background QSOs leaves 26 galaxy and absorber pairs 
and seven galaxies that do not produce corresponding \lya\ absorption lines to
within sensitive upper limits. Redshifts of the galaxy and absorber pairs range
from 0.0750 to 0.8912 with a median of 0.3718, and impact parameter separations
of the galaxy and absorber pairs range from 12.4 to $157.4 \ h^{-1}$ kpc with a
median of $62.4 \ h^{-1}$ kpc.  The primary conclusions are as follows:

  1.  The \lya\ absorption equivalent width $W$ is anti-correlated with galaxy
impact parameter $\rho$ at a level of significance between $3.36 \sigma$ and
$4.45 \sigma$.  We conclude that the amount of gas intercepted along the line 
of sight depends on galaxy impact parameter, which argues that the gas is
physically associated with the individual galaxies.

  2.  The anti-correlation between \lya\ absorption equivalent width $W$ and
galaxy impact parameter $\rho$ accounting for galaxy $B$-band luminosity $L_B$
is superior to the fiducial relationship between $W$ and $\rho$.  We conclude
that the amount of gas intercepted along the line of sight depends on galaxy
$B$-band luminosity.

  3.  The anti-correlation between \lya\ absorption equivalent width $W$ and
galaxy impact parameter $\rho$ accounting for galaxy effective radius $r_e$
is superior to the fiducial relationship between $W$ and $\rho$ but marginally
inferior to the relationship between $W$ and $\rho$ accounting for $L_B$.  We
conclude that the amount of gas intercepted along the line of sight depends on
galaxy effective radius, but only indirectly through the correlation between
galaxy effective radius and galaxy $B$-band luminosity.

  4.  The anti-correlation between \lya\ absorption equivalent width $W$ and
galaxy impact parameter $\rho$ accounting for galaxy average surface brightness
$\langle \mu \rangle$, disk-to-bulge ratio $D/B$, or redshift $z$ is
statistically identical to the fiducial relationship between $W$ and $\rho$.  
We conclude that the amount of gas intercepted along the line of sight does not
depend strongly on galaxy average surface brightness, galaxy disk-to-bulge
ratio, or redshift.  Apparently extended gaseous envelopes are a common and
generic feature of galaxies of a wide range of luminosity and morphological
type, and \lya\ absorption systems trace a significant and representative
portion of the galaxy population. 

  5.  The anti-correlation between \lya\ absorption equivalent width $W$ or $W
\cos i$ and galaxy galactocentric radius $R$ is statistically identical to the
fiducial relationship between $W$ and $\rho$.  We conclude that spherical halos
cannot be distinguished from flattened disks on the basis of the current
observations.

  6.  Most of the galaxies physically associated with \lya\ absorbers do not 
exhibit obvious signs of disturbed morphologies.  We conclude that there is no 
evidence that tenuous gas is distributed around galaxies as a result of galaxy 
interactions in most cases, although it is of course possible that tenuous gas 
is distributed around galaxies as a result of galaxy interactions in some 
cases.

  7.  Incorporating significant scaling relationships (between $W$ and $\rho$
and $L_B$) and rejecting insignificant scaling relationships, our current best
estimate of the column density distribution of tenuous gas around galaxies is
\begin{equation}
\log \left( \frac{N}{10^{20} \ \cmjj} \right)  = -5.33 \; \log \left(
\frac{\rho}{10 \ {\rm kpc}} \right) - 2.19 \; \log \left( \frac{L_B}{L_{B_*}}
\right) + 1.09.
\end{equation}
The cosmic scatter about this relationship is
\begin{equation}
\sigma_c = 0.815,
\end{equation}
which corresponds to a factor of $\approx 6.5$ in neutral hydrogen column
density.

  8.  The relationship between galaxy gaseous radius $r$ and galaxy $B$-band
luminosity $L_B$, which is analogous to the Holmberg (1975) relationship 
between galaxy stellar radius and galaxy $B$-band luminosity, can be described 
by
\begin{equation}
\frac{r}{r_*} = \left( \frac{L_B}{L_{B_*}} \right)^t,
\end{equation}
with
\begin{equation}
t = 0.37 \pm 0.11
\end{equation}
and 
\begin{equation}
r_* = 174_{-25}^{+38} \ h^{-1} \ {\rm kpc}
\end{equation}
for a \lya\ absorption equivalent width threshold $W = 0.3$ \AA. It is clear
that $t=0$ (no dependence of gaseous radius on galaxy luminosity) can be ruled
out at the $3.4 \sigma$ significance level.

  9.  Applying the relationship between galaxy gaseous radius and galaxy
$B$-band luminosity to predict the incidence of \lya\ absorption systems
produced by galaxies drawn from a known galaxy luminosity function, we find
that luminous galaxies can explain about 50\% of \lya\ absorption systems
with $W > 0.3$ \AA.  By including lower luminosity galaxies, we find that 
galaxies might account for all \lya\ absorption systems with $W > 0.3$ \AA. 
But this depends on the unknown luminosity function and gaseous cross sections
of low-luminosity galaxies as well as on the uncertainties of the observed 
number density of \lya\ absorption systems.

\acknowledgments

  The authors thank the staff of STScI for their expert assistance.  H.-W.C. 
and K.M.L. were supported by grant NAG--53261 from NASA; grants 
GO--0594--80194A, GO--0594--90194A, and GO--0661--20195A from STScI; and grant 
AST--9624216 from NSF.  X.B. was partially supported by the DGES under project 
PB95--0122 and acknowledges sabbatical support at Cambridge by the DGES under 
grant PR95---490.

\newpage

\begin{center}
\begin{tabular}{p{1.5in}cccccc}
\multicolumn{7}{c}{TABLE 1} \\
\multicolumn{7}{c}{JOURNAL OF OBSERVATIONS} \\
\tableline
\tableline
& & & & & Exposure \\
\multicolumn{1}{c}{Field} & $\alpha$ & $\delta$ & $z_{\rm em}$ & Filter & Time
(s) & Date \\
\tableline
0349$-$1438 \dotfill & 03:49:09.5 & $-$14:38:07.0 & 0.616 & F702W & 2400 &  5
Nov 1995\\
0405$-$1219 \dotfill & 04:05:27.5 & $-$12:19:31.8 & 0.574 & F702W & 2400 & 18
Jan 1996\\
0850$+$4400 \dotfill & 08:50:13.6 & $+$44:00:29.0 & 0.513 & F702W & 3200 &  7
Feb 1996\\
1001$+$2910 \dotfill & 10:01:10.7 & $+$29:10:09.0 & 0.329 & F702W & 2400 & 24
Nov 1995\\ 
1354$+$1933 \dotfill & 13:54:42.1 & $+$19:33:43.9 & 0.719 & F702W & 2400 & 12
May 1996\\
1704$+$6048 \dotfill & 17:04:03.5 & $+$60:48:31.1 & 0.371 & F702W & 2400 & 29
Nov 1995\\
\tableline
\end{tabular}
\end{center}

\begin{center}
\begin{tabular}{p{1.5in}cccccc}
\multicolumn{7}{c}{TABLE 2} \\
\multicolumn{7}{c}{JOURNAL OF ARCHIVAL OBSERVATIONS} \\
\tableline
\tableline
& & & & & Exposure \\
\multicolumn{1}{c}{Field} & $\alpha$ & $\delta$ & $z_{\rm em}$ & Filter & Time
(s) & Date \\
\tableline
0454$-$2203 \dotfill & 04:54:02.2 & $-$22:03:56.0 & 0.534 & F702W & 1200 &  6
Feb 1994\\
1545$+$2101 \dotfill & 15:45:31.1 & $+$21:01:27.5 & 0.264 & F606W & 1800 &  9
Jun 1994\\
1622$+$2352 \dotfill & 16:22:32.2 & $+$23:52:02.0 & 0.927 & F702W & 24000 & 20
Dec 1994\\
2135$-$1446 \dotfill & 21:35:01.2 & $-$14:46:27.3 & 0.200 & F606W & 2100 & 14
Aug 1994\\
\tableline
\end{tabular}
\end{center}

\begin{center}
\begin{tabular}{p{1.5in}ccccc}
\multicolumn{6}{c}{TABLE 3} \\
\multicolumn{6}{c}{SUMMARY OF OTHER OBSERVATIONS} \\
\tableline
\tableline
& \multicolumn{2}{c}{Galaxies} & & \multicolumn{2}{c}{Absorbers} \\
\cline{2-3}
\cline{5-6}
& Number & & & Number \\
\multicolumn{1}{c}{Field} & Included & Reference & &
Included & Reference \\
\tableline
0349$-$1438 \dotfill &  5 &   1 & &  1 &   2,3 \\
0405$-$1219 \dotfill & 13 & 3,4 & &  4 &     3 \\
0454$-$2203 \dotfill &  8 &   3 & &  3 &     3 \\
0850$+$4400 \dotfill &  7 & 3,5 & &  1 & 2,3,5 \\
1001$+$2910 \dotfill &  2 &   1 & &  2 &   2,3 \\
1354$+$1933 \dotfill &  5 &   1 & &  3 &     3 \\
1545$+$2101 \dotfill &  4 &   6 & &  2 &   3,6 \\
1622$+$2352 \dotfill &  9 &   7 & &  8 &   3,7 \\
1704$+$6048 \dotfill &  5 & 1,8 & &  4 &     3 \\
2135$-$1446 \dotfill &  7 &   3 & &  3 &     3 \\
\tableline
\end{tabular}
\parbox{4.75 in}{\hspace{0.25 in} REFERENCES---(1) Lanzetta et al.\ 1995; (2)
Bahcall et al.\ 1993; (3) our own observations and analysis, in preparation; 
(4) Ellingson \& Yee 1994; (5) Lanzetta et al.\ 1997c; (6) Lanzetta et al.\ 
1996; (7) Steidel et al.\ 1997; (8) Le Brun, Bergeron, \& Boiss\'e 1996.} 
\end{center}

\newpage

\setcounter{page}{31}

\newpage

\figcaption{Final images of galaxies obtained with HST using WFPC2 with the
F606W (for the 1545$+$2101 and 2135$-$1446 fields) or F702W (for the other
fields) filter.  The spatial extent of each image is roughly $25 \ h^{-1}$ kpc 
on a side, and orientation of each image is arbitrary.}

\figcaption{Logarithm of \lya\ rest-frame equivalent width $W$ vs.\ logarithm of
galaxy impact parameter $\rho$.  Circles represent early-type elliptical or S0 
galaxies, triangles represent early-type spiral galaxies, and squares represent
late-type spiral galaxies; small open symbols represent galaxies of luminosity
$L_B < 0.1 L_{B_*}$, small filled symbols  represent galaxies of luminosity $0.1
L_{B_*} < L_B < L_{B_*}$, and large  filled symbols represent galaxies of
luminosity $L_B > L_{B_*}$; and symbols with arrows indicate $3 \sigma$ upper
limits.  The cosmic scatter is indicated by the error bar in the upper-right
corner.}

\figcaption{Logarithm of \lya\ rest-frame equivalent width $W$ vs.\ logarithm
of galaxy impact parameter $\rho$ scaled by galaxy $B$-band luminosity.  The
scaling factor is determined from the analysis described in \S\ 6.  Symbols are
as in Figure 2, and the cosmic scatter is indicated by the error bar in the
upper-right corner.}

\figcaption{Logarithm of \lya\ rest-frame equivalent width $W$ times cosine
of galaxy inclinations angle $i$ vs.\ logarithm of galaxy galactocentric radius
$R$.  Symbols are as in Figure 2, and the cosmic scatter is indicated by the
error bar in the upper-right corner.}

\figcaption{Logarithm of neutral hydrogen column density $N$ vs.\ logarithm
of galaxy impact parameter $\rho$ scaled by galaxy $B$-band luminosity.  The
scaling factor is determined from the analysis described in \S\ 6.  Neutral
hydrogen column densities are determined from \lya\ rest-frame equivalent
widths under the assumption that Doppler parameters are contained in the range
$20 < b < 40$ \kms.  Symbols are as in Figure 2, and the cosmic scatter is
indicated by the error bar in the upper-right corner.}

\end{document}